\documentclass[structabstract]{aa}
\usepackage{graphicx}
\usepackage{txfonts}
\begin{document}
   \title{Near-infrared follow-up to the May 2008 activation of SGR 1627-41\thanks{Based on observations made with ESO VLT at Paranal Observatory under programme ID 281.D-5019}}

   \author{A.~de~Ugarte~Postigo
          \inst{1}
          \and
          A.J.~Castro-Tirado \inst{2}
          \and
          S.~Covino \inst{3}
          \and
          J.~Gorosabel \inst{2}
          \and
          P.~D'Avanzo \inst{3}
          \and
          D.E.A. ~N\"urnberger \inst{1}
          }

   \institute{European Southern Observatory, Casilla 19001, Santiago 19, Chile.\\
              \email{adeugart@eso.org}
         \and
             Instituto de Astrof\' isica de Andaluc\' ia (IAA-CSIC), Camino Bajo de Hu\'etor, 50, E18008, Granada, Spain.
         \and
             INAF - Osservatorio Astronomico di Brera, via Bianchi 46, I-23807 Merate, Italy.
             }

   \date{Received; accepted}

 
  \abstract
   {On 28 May 2008, the \textit{Swift} satellite detected the first reactivation of SGR 1627-41 since its discovery in 1998.}
   {Following this event we began an observing campaign in near infrared wavelengths to search for a possible counterpart inside the error circle of this SGR, which is expected to show flaring activity simultaneous to the high energy flares or at least some variability as compared to the quiescent state.}
   {For the follow-up we used the 0.6m REM robotic telescope at La Silla Observatory, which allowed a fast response within 24 hours and, through director discretionary time, the 8.2m Very Large Telescope at Paranal Observatory. There, we observed with NACO to produce high angular resolution imaging with the aid of adaptive optics.}
   {These observations represent the fastest near infrared observations after an activation of this SGR and the deepest and highest spatial resolution observations of the \textit{Chandra} error circle.}
   {5 sources are detected in the immediate vicinity of the most precise X-ray localisation of this source. For 4 of them we do not detect variability, although the X-ray counterpart experimented a significant decay during our observation period. The 5th source is only detected in one epoch, where we have the best image quality, so no variability constrains can be imposed and remains as the only plausible counterpart. We can impose a limit of $Ks > $ 21.6 magnitudes to any other counterpart candidate one week after the onset of the activity. Our adaptive optics imaging, with a resolution of 0.2$^{\prime\prime}$ provides a reference frame for subsequent studies of future periods of activity.}


   \maketitle
%

\section{Introduction}

   Soft Gamma-ray Repeaters (SGR) are a rare type of astronomical sources that 
   experiment periods of activity in which they show strong flares
   in high energy bands (X-rays and $\gamma$-rays) with typical 
   durations of 0.01 - 1s (\cite{mer08}). These periods of 
   activity usually last between few days and few weeks and are
   separated by periods of quiescence of several years, where a much
   weaker persistent X-ray source is generally observed. Identifications of optical 
   and near infrared (nIR) counterparts, crucial for the precise 
   localisation and broad-band spectral characterisation of these sources are 
   beginning to be achieved (\cite{isr05,kos05,tan08,fat08}) but are still very uncommon.

   Since 1979 and to the end of October 2008 only 5 SGRs have been 
   identified. Four of them lie on the Galactic Plane (SGR 1627-41,
   SGR 1806-20, SGR 1900+14 and the recently discovered SGR 0501+4516)
   while a fifth one is hosted by the Large Magellanic Cloud (SGR 0526-66).
   A parallel family of sources, the Anomalous X-ray Pulsars (AXP) display
   similar behavior, with persistent emission and outburst periods.
   However, in this case, their burst activity is usually limited to the less energetic X-rays. There is much
   discussion regarding the differences and similarities of these two 
   types of events. In 2002, AXP 1E 2259+586 showed a period of 
   activity in which it produced over 80 burst with similar properties to 
   those observed in SGRs, as described by \cite{woo04} and \cite{gav04}. More recently, during the latest 
   activity period of AXP 1E 1547.0-5408 in October 2008 (\cite{kri08a}), 
   it behaved more like a SGR (\cite{kri08b}), being its bursts detected 
   in the $\gamma$-ray range. These examples provide strong evidences of the link between the two families.
   A further source, SWIFT J195509+261406, discovered in June 2007 \cite{pag07,ste08}, 
   showed a similar behavior to SGRs, with an initial $\gamma$-ray spike and 
   having the bulk of the flares detected at X-ray and optical wavelengths (\cite{cas08,ste08,kas08}). 
   It has been suggested to be a magnetar linking SGRs/AXPs and dim isolated neutron 
   stars (\cite{cas08}). The extensive observing campaign performed for this source
   in optical and nIR is a good reference to compare with the behavior of the 
   counterparts of these kind of sources.

   The powering sources of SGRs and AXPs are thought to be strongly magnetised
   neutron stars or ``magnetars'' \cite{dun92}, that shine in high energies due 
   to the decay of the magnetic fields. In this model, neutron stars with initial
   fields of $\sim10^{14}-10^{15}$ G would be observed as X-ray pulsars with a 
   spin down due to magnetic dipole radiation, consistent with what has been observed
   in SGRs. The relation of magnetars with neutron stars is supported by the fact that 
   there have been several associations between AXPs and supernova remnants, that 
   would be the result of the explosions that generated the neutron stars 
   (\cite{gae01,mer08}). On the other hand, several SGRs are likely associated with clusters of 
   massive stars with the potential to host objects that evolve to form 
   the required neutron stars (\cite{mer08,wac08}).

   In this paper we report the follow-up to the second registered activation of SGR 1627-41,
   which began on 28 May 2008 \cite{pal08}, after a ten year period of quiescence.
   This SGR was first detected on 15 June 1998 \cite{kou98}, when it began a period of 
   activity of 6 weeks, during which it produced of the order of 100 bursts \cite{woo99}.
   We used this new opportunity to obtain deep nIR imaging, trying to identify 
   the counterpart to this event. SGR 1627-41 has the peculiarity of being the only known
   SGR with a transient behavior, with a persistent X-ray emission that 
   decayed monotonically with time between the activity period of 1998 and the
   one of 2008 \cite{mer06}. The X-ray emission increase after this new activation has been 
   used by \cite{esp09} to determine a pulsation period of 2.6 seconds for this source, unknow until now.

\begin{table*}
\caption{Observing log of the follow-up campaign for the May 2008 activation of SGR 1627-41.}             
\label{table:1}      
\centering                          
\begin{tabular}{c c c c c c c}        
\hline\hline                 
Observation Interval   & Time since     & Telescope    & Filter             & Exposure     & FWHM                 & Limiting \\  
(2008 UT)              & trigger (days) &+ Instrument &                    & (s)          &                      & Magnitude \\ 
\hline                        
29.3411 - 29.3803 May  & 1.012         & REM+REMIR    & \textit{Ks}        & 50$\times$12 & 3.6$^{\prime\prime}$ & 14.5 \\
29.3424 - 29.3817 May  & 1.014         & REM+REMIR    & \textit{J}         & 50$\times$12 & 4.0$^{\prime\prime}$ & 16.5 \\
29.3437 - 29.3830 May  & 1.015         & REM+REMIR    & \textit{H}         & 60$\times$12 & 3.8$^{\prime\prime}$ & 15.4 \\
31.3858 - 31.3961 May  & 3.042         & REM+REMIR    & \textit{H}         & 50$\times$12 & 4.2$^{\prime\prime}$ & 15.2 \\
4.0547 - 4.0953 June   & 6.727         & VLT+NACO     & \textit{Ks}        & 51$\times$60 & 0.2$^{\prime\prime}$ & 21.6 \\
7.0399 - 7.1348 June   & 9.739         & VLT+NACO     & \textit{Ks}        & 114$\times$60& 0.3$^{\prime\prime}$ & 21.3 \\
15.1645 - 15.1747 June & 17.821        & REM+REMIR    & \textit{H}         & 50$\times$12 & 3.7$^{\prime\prime}$ & 15.6 \\
2.9908 - 3.0311 August & 66.662        & VLT+NACO     &\textit{Br$\gamma$}&18$\times$180 & 0.3$^{\prime\prime}$ & ---  \\
3.0344 - 3.0422 August & 66.690        & VLT+NACO     & \textit{Ks}        & 30$\times$60 & 0.4$^{\prime\prime}$ & 20.2 \\
\hline  \hline                                 
\end{tabular}
\end{table*}

   In Sect. 2 we describe the follow-up campaign to SGR 1627-41 as well as the reduction and analysis techniques that were used. Sect. 3 presents the results of the observations and Sect. 4 discusses their implications. Sect. 5 summarises our conclusions. 


\section{Observations}

At 08:21:43 UT of 28 May 2008, the BAT $\gamma$-ray detector aboard the \textit{Swift} mission 
\cite{geh04} was triggered by a bright $\gamma$-ray event that was related to the known 
SGR 1627-41 \cite{pal08}. It was the beginning of an activity period that persisted with tens of other bursts during the following hours (\cite{esp08}).

Following this detection, we triggered observations with the 0.6m REM robotic telescope \cite{chi03} at La Silla Observatory (Chile). Due to bad weather conditions at the observatory, the first images were not obtained until 1.0 days after the first detected burst. Due to the extreme extinction of $A_V\sim54$ magnitudes and although the telescope carries both optical and infrared detectors, only infrared images were taken, as optical observations would impose very limited constrains. The dataset comprises \textit{J}, \textit{H} and \textit{Ks}-band imaging. The data of 29 May were taken in cycles, alternating \textit{Ks}, \textit{J} and \textit{H}-band filters and which were combined to produce one single image in each band.

In parallel, we applied for director discretionary time to obtain deep nIR observations with one of the 8.2m units of the VLT telescope at Paranal Observatory. The observations were performed using NACO on Yepun (the fourth unit telescope of the VLT). NACO (\cite{len03,rou03}) is composed of the adaptive optics module (NAOS) plus the high resolution nIR camera (CONICA). Adaptive optics correction was performed using a nearby natural guide star. We obtained two \textit{Ks}-band epochs plus a late observation in Br$\gamma$ and \textit{Ks}-band, all of them using the S54 camera, which provides a field of view of $56^{\prime\prime}\times56^{\prime\prime}$. Table 1 displays the observing log, including angular resolution and 3-$\sigma$ limiting magnitudes of the combined frames. The exposures are indicated by the number of exposures times the total exposure time per dither possition, which in some cases is a coaddition of shorter exposures in order to avoid saturation.

The reduction was performed using IRAF (\cite{tod93}) through the following steps: First we created a master flat field by combining lamp flats subtracted of dark current. Once normalised, we used this flat to correct all the science images. In the next step, we combined these science frames, which had been observed with offsets between them, to create sky frames. These sky frames were normalised to the background of each science frame and subtracted to create the final reduced science images, corrected of the sky contribution. In the last step these reduced science frames were aligned and combined together to create a deep image for each epoch, as displayed in Table 1.

   \begin{figure}
   \centering
   \includegraphics[width=9cm]{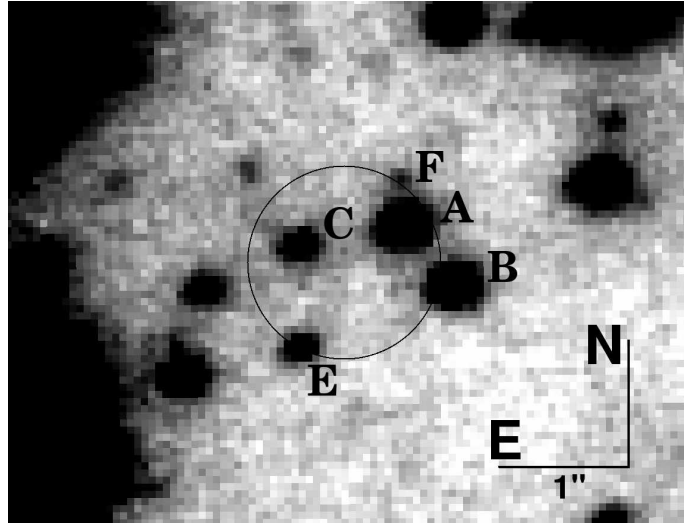}
   \caption{NACO image showing the \textit{Chandra} error circle of SGR 1627-41 on 4 June 2008. The error box has taken into account both the uncertainty of the \textit{Chandra} coordinates and the precision of our astrometry (3-$\sigma$). The field of view is $5.7^{\prime\prime}\times3.2^{\prime\prime}$.}
    \label{FigNACO}%
    \end{figure}

\begin{table}
\caption{Astrometry of the selected objects based on the 2MASS catalogue, with a 1-$\sigma$ uncertainty of $\pm0.14^{\prime\prime}$.}             
\label{table:2}      
\centering                          
\begin{tabular}{c c c}        
\hline\hline                 
Object  & R.A.(J2000)       & Dec.(J2000)  \\    
\hline                        
A	& 16:35:51.789	    & -47:35:23.06 \\
B	& 16:35:51.752	    & -47:35:23.51 \\
C	& 16:35:51.866	    & -47:35:23.20 \\
E	& 16:35:51.868      & -47:35:23.98 \\
F	& 16:35:51.792	    & -47:35:22.70 \\
\hline \hline                                  
\end{tabular}
\end{table}

\begin{table}
\caption{\textit{Ks}-band photometry of the selected objects, using 2MASS stars as reference.}             
\label{table:3}      
\centering                          
\begin{tabular}{c c c c}        
\hline\hline                 
Object &                       &      Date (2008)      &            \\    
       &   4.08 Jun             &  7.09 Jun              &  3.04 Aug   \\ 
\hline                        
A      & 18.08$\pm$0.11 & 18.05$\pm$0.12 & 18.13$\pm$0.16 \\
B      & 18.18$\pm$0.11 & 18.16$\pm$0.12 & 18.42$\pm$0.16 \\
C      & 19.28$\pm$0.12 & 19.32$\pm$0.13 & 19.12$\pm$0.19 \\
E      & 19.57$\pm$0.12 & 19.50$\pm$0.13 & 19.65$\pm$0.20 \\
F      & 20.10$\pm$0.14 & ---            & ---            \\
\hline\hline                                   
\end{tabular}
\end{table}

The astrometry of each of the images was obtained using JIBARO codes \cite{deu05}, taking as reference all the objects of the 2MASS all-sky catalogue of point sources \cite{cut03} present in each of the frames, while avoiding any saturated sources. Fig.~1 shows the central region of the NACO image obtained on 4 June 2008. In it we mark a set of selected objects together with the SGR 1627-41 error circle ($Chandra$ 3-$\sigma$ error combined with our 3-$\sigma$ astrometric accuracy). The coordinates given in Table 2 were obtained using the first NACO epoch and correlating with 23 reference objects. We performed aperture photometry using PHOT with IRAF, taking again as reference the 2MASS catalogue. Photometric values for the selected sources are given in Table 3.

\section{Results}

The REM observations presented here are the first fast-response nIR observations after an activity period of SGR 1627-41, starting only 24 hours after the first burst. However, the last reported high energy burst of this activation happened less than one hour after the initial \textit{Swift} trigger, so that none of our observations are simultaneous to any of the bursts. In our data we do not detect any nIR flares in the individual images down to typical 3-$\sigma$ limiting magnitudes of $J\sim$ 15.2, $H\sim$ 14.5, $Ks\sim$ 13.7 for the REM telescope and $Ks\sim$ 20.0 for VLT/NACO. However, we must note that these values are calculated for the integrated exposure times, while the durations of the flares observed for this outburst (\cite{esp08}) were significantly shorter than the exposure times ($\sim 0.1$ s flares vs. 12 s exposures in the case of REM).

Using the deep, high angular resolution NACO data, we identify 5 sources within the immediate vicinity of the  most precise $Chandra$ localisation of SGR 1627-41 (\cite{wac04}, W04 hereafter) which we will refer to as A, B, C, E and F, following the notation of W04 (see Fig. 1). A, B and C were already identified by W04. We do not detect source D of W04 which can be most probably considered just a noise spike in their data, as already suggested by them. Source F can only be clearly separated from source A in the first epoch, where we have the best angular resolution. In the datasets of the remaining epochs they are blended together. In Table 2 we give the photometry of each of these sources in the 3 NACO epochs. The objects are not detected in the REM data.

During the period between the 4th of June and the 3rd of August 2008 the X-ray emission decayed as a powerlaw with an index of $\sim$ -0.2 (\cite{esp08}), implying a flux decrease of a factor $\sim$ 0.6. If the nIR sources were to decay in a similar way we would expect a loss of $\sim$ 0.5 mag during this time. This was the case of other sources such as SGR 1806-20 or SWIFT J195509+261406, where the X-ray and the optical emission were seen to vary in a similar way (\cite{isr05,cas08}). However, all the nIR sources are consistent with no decay between the first and last NACO epochs. Decay limits between them can be imposed as of $\Delta Ks(A) < 0.3$, $\Delta Ks(B) < 0.5$, $\Delta Ks(C) < 0.2$ and $\Delta Ks(E) < 0.4$ (1-$\sigma$). A decay limit can not be imposed for source F due to contamination by source A. Furthermore, we measure no decay of sources A and B as compared to the observations performed by W04 in March 2001, when the X-ray emission was lower by a factor of $\sim$ 10 (2.5 magnitudes). As sources C and E can still be seen in the observations of W04 (Fig. 2 of their paper) we may conclude that the variability of sources A, B, C or E does not correlate with that of the X-ray emission.

On our last NACO visit on 3 August 2008 we obtained narrow band imaging with a 0.023 $\mu$m wide Br$\gamma$ filter centered at 2.166 $\mu$m, searching for H I Br$\gamma$ emission in the spectrum of any of the sources, arising from an accretion disk if the magnetar would be part of a binary system, as seen in some Galactic X-ray binaries (\cite{cas96}). In order to detect any excess in this band we measured the flux of each of the selected objects in Br$\gamma$ and divided it by its flux in $Ks$. We normalised these ratios by dividing them with the average of the same ratio for all the objects in the field, so that a value of 1.0 would imply no Br$\gamma$ excess. The result for each of our selected objects (except for F which is not detected in Br$\gamma$) is displayed in Table 4. None of the objects show any significant excess.

\begin{table}
\caption{Relative fluxes between Br$\gamma$ and $Ks$-band, normalised to the median ratio of all the objects in the field.}             
\label{table:4}      
\centering                          
\begin{tabular}{c c}        
\hline\hline                 
Object &  F$_{Br\gamma}$/F$_{Ks}$            \\    
\hline                        
A      & 1.02$\pm$0.07 \\
B      & 1.12$\pm$0.08 \\
C      & 0.78$\pm$0.09 \\
E      & 0.82$\pm$0.10 \\
F      & --- \\
\hline\hline                                   
\end{tabular}
\end{table}

\section{Discussion}

The extinction at the position of the SGR can be derived from the different X-ray spectra (\cite{esp08}) as A$_V = 54\pm6$ magnitudes. Due to this large extinction, follow-up in optical bands imposes almost no constrains, thus the need for nIR observations. Even so, this implies an extinction in $K$-band of A$_K = 6.0\pm0.7$ magnitudes. This is a common problem for most SGR/AXP follow-ups in optical/nIR bands, as they are generally obscured by the dust in the disk of the Galaxy, although SGR 1627-41 is up to now the case with the most extreme extinction.

During the active period of 1998 optical observations failed to detect any new source down to a limiting magnitude of $Ic\sim$20 (\cite{cas00}). To our knowledge, no nIR follow-up was performed during that activation. Some time later, in March 2001, W04 obtained deep nIR observations of the quiescent source, identifying 3 candidates (A, B and C) within their very precise localisation of the X-ray counterpart that they obtained using \textit{Chandra} data. In the deep and high angular resolution images obtained with VLT/NACO shortly after the May 2008 activation, we again identify these 3 sources. A fourth source (D) that was identified with a low significance by W04 is not seen in our observations and was most probably just a noise spike in their image. We study 2 more sources (E and F) that within the \textit{Chandra} error circle.

Source F is only identified in the first NACO epoch, as it is blended with A in the rest of the frames. In the period between 7 days and 67 days after the outburst onset, no significant variability is detected in any of the other objects, in spite of the factor $\sim$ 0.6 decrease in the X-ray flux. Furthermore, our measurements of sources A and B are consistent with the values of 2001 given by W04, period in which there is an X-ray increase of a factor $\sim$ 10. This is consistent with the suggestion of W04 that neither A or B are probably the counterparts of SGR 1627-41, based on their colour indexes. All these evidences argue against the identification of A, B, C or E as the nIR counterpart to SGR 1627-41, leaving F as the only plausible counterpart candidate. We can impose a 3-$\sigma$ limiting magnitude of $Ks > $21.6 to any other counterpart to the SGR within the \textit{Chandra} error circle. A combination of the data of different epochs does not produce a significant improvement as compared to the frame created with only the first epoch, as the gain in depth due to longer exposure comes together with a significant decrease in resolution.

Up to now, only two counterparts to SGRs have been established. Monitoring of SGR 1806-20 during the 2004 year long activity period, which ended with a giant flare on 27 December 2004, showed a nIR counterpart with variability in the range $Ks\sim$ 18.3 - 21.0 within a 7 month period (\cite{kos05,isr05}). The line of sight extinction to SGR 1806-20 is of A$_K\sim$ 3.5, implying that with the extinction of SGR 1627-41 it would have had a peak magnitude of $Ks\sim$20.8, consistent with the measurement for object F. 

The second accepted counterpart for a SGR is the one of SGR 0501+4516. On 22 August 2008 Tanvir \& Varricatt (2008) responded to the activation (and discovery) of SGR 0501+4516 (\cite{hol08,bar08}) identifying a counterpart of $Ks\sim$ 18.6 starting 2 hours after the first gamma-ray burst. The counterpart decayed during the next days to $Ks\sim$ 19.2 (Rea et al. 2008, de Ugarte Postigo et al. 2008). This counterpart was also identified at optical wavelengths with a magnitude of $Ic \sim$ 23.3 (\cite{fat08,ofe08}). In contrast to SGR 1627-41, SGR 0501+4516 has a very low line of sight extinction, with an A$_K\sim$ 0.2 magnitudes. Obscured by the extinction of the line of sight towards SGR 1627-41 this object would have had a brightest magnitude of $Ks \sim$ 24.5 (we are not considering differences in distances, as the distance to SGR 0501+4516 has not been yet established and the difference would probably not be significant). This would be in agreement with a non-detection of the counterpart in our data.

From the study the of the well linked counterparts to several AXPs, unextincted magnitudes of $Ks$ = 19 - 21 (W04, \cite{isr02,isr03,isr04}) can be derived for most counterparts. This values are very similar to what was obtained for SGR 0501+4516 and also consistent with a non detection in our data.

With these results we may conclude that of the 5 objects that we have studied, 4 of them (A, B, C and E) are very unlikely related with SGR 1627-41 while object F remains as a counterpart candidate. Further high resolution observations during quiescence will be required in order to determine any possible relation of this source with SGR 1627-41.


\section{Conclusions}

\begin{enumerate}

\item We have performed a follow-up campaign to the second registered activation of SGR 1627-41. Our observations began one day after the first burst detected by \textit{Swift}.

\item The May 2008 outburst was extremely short, with no bursts reported after one hour, as compared with the previous activation in 1998, when bursts were produced during 6 weeks. This prevented us from obtaining observations simultaneously to the bursting activity.

\item We study 5 sources within the \textit{Chandra} error circle. There is no evidence of variability of sources A and B as compared to the measurements carried out by W04 in 2001, during quiescence. No variability is observed in objects A, B, C and E within the time range of our observations, in contrast with the X-ray decay. Observations in Br$\gamma$ do not show any evidence of emission signatures in any of the sources.

\item Object F is only detected in the first epoch, where the image quality is best, at a magnitude of $Ks$ = 20.10$\pm$0.14. Thus, we can not impose limits on its possible flux decay. It remains as the only plausible counterpart. Further observations will be required to confirm or discard its relation with SGR 1627-41.

\item The imaging obtained with VLT+NACO, with a best angular resolution of 0.2$^{\prime\prime}$ and limiting magnitude of $Ks\sim$ 21.6 will serve as reference for future observation campaigns.

\item In order to obtain detections in future activation periods, the use of large telescopes with infrared instrumentation is mandatory. Follow-up should ideally be done through the use of ToO programmes, as a fast response is also necessary in order to secure the observation during the activity period.

\end{enumerate}

\begin{acknowledgements}
We are grateful to ESO for granting DDT programme 281.D-5019 at the VLT. AdUP acknowledges support from an ESO fellowship. This research has been partially supported by the Spanish Ministry  of Science through  the programmes ESP2005-07714-C03-03 and AYA 2007-63677. We thank the anonymous referee for constructive comments.
\end{acknowledgements}


\begin{thebibliography}{}
\bibitem[Barthelmy et al. 2008]{bar08} Barthelmy, S.~D., et al.\ 2008, GRB Coordinates Network, 8113, 1
\bibitem[Castro-Tirado et al. 1996]{cas96} Castro-Tirado, A.~J., Geballe, T.~R., \& Lund, N.\ 1996, \apjl, 461, L99 
\bibitem[Castro-Tirado et al. 2000]{cas00} Castro-Tirado, A.~J., Lund, N., Pinfield, D., 
\& Covino, S.\ 2000, Gamma-ray Bursts, 5th Huntsville Symposium, 526, 801 
\bibitem[Castro-Tirado et al. 2008]{cas08} Castro-Tirado, A.~J., et al.\ 2008, \nat, 455, 506 
\bibitem[(Chincarini et al. 2003)]{chi03} Chincarini, G., et al.\ 2003, The Messenger, 113, 40 
\bibitem[(Cutri et al. 2003)]{cut03} Cutri, R.~M., et al.\ 2003, The IRSA 2MASS All-Sky Point Source Catalog, NASA/IPAC Infrared Science Archive. http://irsa.ipac.caltech.edu/applications/Gator/
\bibitem[(Duncan \& Thompson 1992)]{dun92} Duncan, R.~C., \& Thompson, C.\ 1992, \apjl, 392, L9 
\bibitem[Esposito et al. 2008]{esp08} Esposito, P., et al.\ 2008, \mnras, 390, L34 
\bibitem[Esposito et al. (2009)]{esp09} Esposito, P., et al.\ 2009, \apjl, 690, L105 
\bibitem[Fatkhullin et al. 2008]{fat08} Fatkhullin, T., et al.\ 2008, GRB Coordinates Network, 8160, 1
\bibitem[Gaensler et al. 2001]{gae01} Gaensler, B.~M., Slane, P.~O., Gotthelf, E.~V., \& Vasisht, G.\ 2001, \apj, 559, 963 
\bibitem[Gavriil et al. (2004)]{gav04} Gavriil, F.~P., Kaspi, V.~M., \& Woods, P.~M.\ 2004, \apj, 607, 959
\bibitem[(Gehrels et al. 2004)]{geh04} Gehrels, N., et al.\ 2004, \apj, 611, 1005 
\bibitem[Holland et al. 2008]{hol08} Holland, S.~T., et al.\ 2008, GRB Coordinates Network, 8112, 1 
\bibitem[Israel et al. 2002]{isr02} Israel, G.~L., et al.\ 2002, \apjl, 580, L143
\bibitem[Israel et al. 2003]{isr03} Israel, G.~L., et al.\ 2003, \apjl, 589, L93
\bibitem[Israel et al. 2004]{isr04} Israel, G.~L., et al.\ 2004, \apjl, 603, L97 
\bibitem[Israel et al. 2005]{isr05} Israel, G., et al.\ 2005, \aap, 438, L1 
\bibitem[Kasliwal et al. 2008]{kas08} Kasliwal, M.~M., et al.\ 2008, \apj, 678, 1127
\bibitem[Kosugi et al. 2005]{kos05} Kosugi, G., Ogasawara, R., \& Terada, H.\ 2005, \apjl, 623, L125
\bibitem[(Kouveliotou et al. 1998)]{kou98} Kouveliotou, C., Kippen, M., Woods, P., Richardson, G., Connaughton, V., \& McCollough, M.\ 1998, \iaucirc, 6944, 2
\bibitem[Krimm et al. 2008a]{kri08a} Krimm, H.~A., et al.\ 2008a, GRB Coordinates Network, 8311, 1
\bibitem[Krimm et al. 2008b]{kri08b} Krimm, H.~A., Beardmore, A.~P., Gehrels, N., Page, K.~L., Palmer, D.~M., Starling, R.~L.~C., \& Ukwatta, T.~N.\ 2008b, GRB Coordinates Network, 8312, 1 
\bibitem[Lenzen et al. 2003]{len03} Lenzen, R., et al.\ 2003, \procspie, 4841, 944 
\bibitem[(Mereghetti et al. 2006)]{mer06} Mereghetti, S., et al.\ 2006, \aap, 450, 759
\bibitem[Mereghetti 2008]{mer08} Mereghetti, S.\ 2008, \aapr, 15, 225
\bibitem[Ofek et al. 2008]{ofe08} Ofek, E.~O., Kiewe, M., \& Arcavi, I.\ 2008, GRB Coordinates Network, 8229, 1
\bibitem[(Pagani et al. 2007]{pag07} Pagani, C., et al.\ 2007, GRB Coordinates Network, 6489, 1 
\bibitem[(Palmer et al. 2008)]{pal08} Palmer, D., et al.\ 2008, GRB Coordinates Network, 7777, 1 
\bibitem[Rea et al. (2008)]{rea08} Rea, N., Rol, E., Curran, P.~A., Skillen, I., Russell, D.~M., \& Israel, G.~L.\ 2008, GRB Coordinates Network, 8159, 1
 \bibitem[Rousset et al. 2003]{rou03} Rousset, G., et al.\ 2003, \procspie, 4839, 140
\bibitem[Stefanescu et al. 2008]{ste08} Stefanescu, A., Kanbach, G., S{\l}owikowska, A., Greiner, J., McBreen, S., \& Sala, G.\ 2008, \nat, 455, 503 
\bibitem[Tanvir \& Varricatt 2008]{tan08} Tanvir, N.~R., \& Varricatt, W.\ 2008, GRB Coordinates Network, 8126, 1
\bibitem[Tody 1993]{tod93} Tody, D.\ 1993, Astronomical Data Analysis Software and Systems II, 52, 173 
\bibitem[(de Ugarte Postigo et al. 2005)]{deu05} de Ugarte Postigo, A., et al.\ 2005, Astrof\' isica Rob\'otica en Espa\~na, edited by A.J. Castro-Tirado, B.A. de la Morena and J. Torres, Madrid, pp. 35-50.
\bibitem[de Ugarte Postigo et al.(2008)]{deu08} de Ugarte Postigo, A., Castro-Tirado, A.~J., Gorosabel, J., Morales-Calderon, M., \& Huelamo, N.\ 2008, GRB Coordinates Network, 8162, 1 
\bibitem[Wachter et al. 2004]{wac04} Wachter, S., et al.\ 2004, \apj, 615, 887 
\bibitem[Wachter et al. 2008]{wac08} Wachter, S., Ramirez-Ruiz, E., Dwarkadas, V.~V., Kouveliotou, C., Granot, J., Patel, S.~K., \& Figer, D.\ 2008, \nat, 453, 626 
\bibitem[(Woods et al. 1999)]{woo99} Woods, P.~M., et al.\ 1999, \apjl, 519, L139
\bibitem[Woods et al. (2004)]{woo04} Woods, P.~M., et al.\ 2004, \apj, 605, 378 



  
\end{thebibliography}
\end{document}